\documentclass[12pt]{article}
\usepackage[T1]{fontenc}
\usepackage[utf8]{inputenc}
\usepackage{authblk}
\usepackage{graphicx}
\usepackage{epstopdf}
\usepackage{amssymb}
\usepackage{cite}
\usepackage{color}
\usepackage[binary,amssymb]{SIunits}
\usepackage{amsfonts}
\usepackage{mathtools}

\newcommand{\x}{\text{x}}

\usepackage{color} 

\title{Multiband superconductivity in ${\rm BiS_2}$-based layered compounds}
\author[1]{M. A. Griffith}
\author[1]{T. O. Puel}
\author[1]{M. A. Continentino}
\author[2]{G. B. Martins}

\affil[1]{Centro Brasileiro de Pesquisas F\'{\i}sicas, Rua Dr. Xavier Sigaud 150, Urca, 22290-180 Rio de Janeiro, RJ, Brazil}
\affil[2]{Instituto de F\'{\i}sica, Universidade Federal Fluminense, 24210-346 Niter\'oi, RJ, Brazil}

\begin{document}
\maketitle
\begin{abstract}
A mean-field treatment is presented of a square lattice two-orbital-model for ${\rm BiS_2}$ 
taking into account intra- and inter-orbital superconductivity. A rich phase diagram involving both 
types of superconductivity is presented as a function of the ratio 
between the couplings of electrons in the same and different orbitals 
(${\rm \eta= V_{XX}/V_{XY}}$) and electron doping $\x$. 
With the help of a quantity we call \emph{orbital-mixing ratio}, denoted as $R(\phi)$, 
the phase diagram is analyzed using a simple and intuitive picture based on 
how $R(\phi)$ varies as electron doping increases. The predictive power of $R(\phi)$ 
suggests that it could be a useful tool in qualitatively (or even semi-quantitatively) 
analyzing multiband superconductivity in BCS-like superconductors. 
\end{abstract}


\section{Introduction}\label{intro} 

The study of two-band superconductivity (SC) \cite{Suhl1959,Annette,Lin2014} (or, more generally, multiband SC) 
has become of increasing relevance as superconducting materials with overlapping bands at the Fermi surface, 
like, for example, ${\rm MgB_2}$ \cite{Buzea2001,Xi2008}, are discovered. 
What distinguishes these systems from the more traditional single-band case is the coexistence, 
at the Fermi level, of electrons from different bands (originating from different orbitals). 
These electrons, which are directly involved 
in the superconducting ground state, can, in principle, pair in a variety of ways. 
The large class of multiband superconductors includes heavy-fermion systems \cite{Seyfarth2005,Jourdan2004}, 
the well-studied ${\rm MgB_2}$ \cite{Floris2007}, the 
pnictides \cite{Thomale2013} and, more recently, the layered sulfides ${\rm BiS_2}$ \cite{Biswas2013}. 
The types of pairing in multiband systems can be categorized 
in two main groups, namely intraband and interband pairing, depending on 
the predominant paring interaction in the system. These two types of pairings are not mutually exclusive, they 
may coexist and even `compete' in the same material, changing in relative importance  as some external 
parameter, as pressure or doping, is varied \cite{Tanaka2015}. The superconducting state resulting from 
the addition of a second band \cite{Suhl1959,Kondo1963} to the traditional single-band BCS 
state \cite{Bardeen1957} shows many interesting new features, like the possibility of formation 
of two superconducting gaps, which may then be observed by either 
Angle-Resolved Photoemission Spectroscopy (ARPES)\cite{Ding2008}, Scanning Tunneling Spectroscopy (STS) \cite{Iavarone2002}, 
or thermal transport measurements under magnetic field \cite{Seyfarth2005}, for example; the possibility of pairing even 
when the electron-electron interaction in one of the bands is repulsive, in which case, when an interaction between 
the bands is introduced, $T_c$ increases in comparison to the single-band attractive case\cite{Suhl1959}. 
In addition, the isotope effect vanishes when the interband interaction is large, explaining 
the behavior of superconductors like ${\rm Nb_3Sn}$ \cite{Kondo1963}. 
Note that the motivation for Suhl {\it et al.} \cite{Suhl1959} to introduce the `extra' band was to try 
and explain the (relatively) high-${\rm T_c}$ observed in transition metal superconducting compounds \cite{Gladstone}. 
An indication that a second band had to be taken into account to treat SC in the transition metal elements 
was that s-d electron scattering seemed to be important to explain their resistivity in 
the normal state. Suhl {\it et al.} \cite{Suhl1959} 
analyzed three different situations (denoting the intraband pairing interactions as ${\rm V_{ss}}$ and ${\rm V_{dd}}$, 
and the interband as ${\rm V_{sd}}$): (i) finite ${\rm V_{sd}}$ and ${\rm V_{ss}=V_{dd}=0}$, obtaining two different gaps 
(unless the density of states $\rho_s=\rho_d$, in which case the gaps are equal) whose dependence on temperature 
is BCS-like, but that, nonetheless, have the same $T_c$; (ii) ${\rm V_{sd}=0}$, where there are two gaps as well, 
with a BCS-like temperature dependence, however, with two different $T_c$ values; and (iii) if 
a small ${\rm V_{sd} \ll \sqrt{V_{ss}V_{dd}}}$ is turned on, a single $T_c$ is obtained that is close 
but always above the larger $T_c$ 
in (ii), as well as a gap with a dependence on temperature that is an interpolation between the gaps obtained in (ii). 
Two clear examples of case (i) can be observed first in ${\rm MgB_2}$ 
trough STS data as a function of temperature \cite{Iavarone2002}, 
and second in the pnictide compound ${\rm Ba_{0.6}K_{0.4}Fe_2As_2}$ 
through ARPES \cite{Ding2008}.  

In this work, using an orbital basis, we will study the contribution of different types of 
pairings, intra- and inter-orbital, 
to the superconducting phase of ${\rm BiS_2}$ systems as a function of electron doping.  
Reference \cite{Hirsch2015}, where 32 classes of superconductors 
were studied, placed the ${\rm BiS_2}$ family of superconductors in the `possibly 
unconventional' column. Reference \cite{Yazici2015} describes the experiments that  
suggest the possibility of these materials exhibiting unconventional superconductivity. 
The fact that these results come from polycrystalline samples, which are prone to inhomogeneities 
and random orientation of crystallites (which becomes relevant for measurements depending on 
the application of a magnetic field) warrants the cautious approach taken by the community 
working on ${\rm BiS_2}$. Thus, in the present work, we do not assume any specific pairing mechanism, 
although we briefly refer to `phonon pair-scattering', for the sake of argument, when discussing the 
results.  As to the SC gap, muon-spin rotation ($\mu$SR) experiments \cite{Biswas2013}, for example, support 
multiband SC in the ${\rm BiS_2}$-based layered compound ${\rm Bi_4O_4S_3}$, pointing 
to two s-wave-type energy gaps, although the authors 
do not rule out the possibility of fitting the data with a single s-wave gap.
Therefore, we will consider these materials as s-wave ($\vec{k}$-independent) singlet superconductors 
and use a mean-field approach to analyze its double-gap properties. 
In general, interband pairing between bands which cross the 
Fermi surface at different wave-vectors may favor the appearance 
of inhomogeneous superconducting states characterized by a  wave-vector 
Q corresponding to the difference between the different band wave-vectors \cite{Fulde1964,Larkin1965,Padilha2009}.  
No evidence of such phenomenon has been experimentally observed in ${\rm BiS_2}$ compounds, therefore we do not 
take this possibility into account in our model. 
Features associated with low dimensionality \cite{Usui2012a,Sugimoto2015a,Griffith2016} are important to determine 
the electronic structure of these materials, but, close to the SC transition, fluctuations are averaged out 
as indicated by the large coherence length measured for these materials \cite{Awana2013,Srivastava2014}, 
thus justifying the use of a BCS (mean-field) treatment of the problem, as undertaken here. 
Aside from the controversy regarding the pairing mechanism, superconductors based on ${\rm BiS_2}$ layers 
have revealed complex and surprising properties. For example, recently, coexistence of magnetism and SC has 
been reported~\cite{Feng2016} in ${\rm Bi_{4-x}Mn_{x}O_4S_3}$. These phenomena  are observed in different layers  of the 
system and appear as rather independent of each other. The substitution by Co and Ni instead of Mn suggests 
that the increase in $T_c$ due to the latter can be attributed to its mixed valence, which allows for an 
effective charge transfer to the superconducting layers.

This work is divided as follows: In section \ref{tight} we present the tight-binding two-orbital model for ${\rm BiS_2}$, 
showing in detail how does the Fermi surface changes with electron doping. 
Section \ref{inth} presents the pairing interactions we are considering, while section \ref{gapeq} develops 
the gap equations at the 
mean-field level. Section \ref{symm} closes with some simplifying assumptions regarding the paring interactions, which reduce the 
number of gap equations from 4 to 2. We close section II by presenting the solution to the gap equations as a 
function of ${\rm \eta=V_{XX}/V_{XY}}$ (the ratio of the relevant pairing couplings) and the electron doping $\x$. 
In Section \ref{ratiosec}, we clearly define what is meant by orbital-mixing, by introducing the quantity $R(\phi)$ 
to measure it along the Fermi surface, and describe its relevance to multiband SC. Section \ref{low} 
describes how the structures seen in both gap functions below the Lifshitz transition can be understood 
through the way $R(\phi)$ changes with doping. In section \ref{high} the same is done above the Lifshitz transition. 
In addition, section \ref{tc} presents results for the superconducting critical temperature $T_c$, which 
are qualitatively in agreement with those for ${\rm BiS_2}$ compounds. 
The paper closes with section \ref{conc}, where Summary and Conclusions are given. 

\begin{table}
\caption{Tight-binding parameters (eV) for two-orbital model.\label{tab:hopp3}}
 \begin{tabular}{|ccccccccc|}\hline
$\epsilon_{X,Y}$ & $t_x^{X,Y}$ & $t_{x \mp y}^{X,Y}$ & $t_{x \pm y}^{X,Y}$ & $t_{2x \mp y}^{X,Y}$ & $t_{2x \pm y}^{X,Y}$ & $t_x^{XY}$ & $t_{2x}^{XY}$ & $t_{2x+y}^{XY}$\\
\hline
  $2.811$   & $-0.167$ & $0.880$  & $0.094$ & $0.069$ & $0.014$ & $0.107$ & $-0.028$ & $0.020$ \\ 
\hline
 \end{tabular}
\end{table}

The main message of this work is that the systematic application of the orbital-mixing concept to systems showing 
BCS-like multiband SC can pinpoint regions of the phase diagram where one of the possible superconducting order 
parameters may dominate over the others, or where, for example, a competition between different order parameters may occur. 
The concept is illustrated 
through its detailed application to a two-orbital model for ${\rm BiS_2}$, which, due to the marked dependence 
of its Fermi surface on electron doping and the well defined variation of orbital-mixing along the BZ, 
provides a particularly convincing connection between orbital-mixing and specific superconducting order parameters. 

\section{Model}

\subsection{Tight-binding two-orbital model}\label{tight}

The  electronic structure of ${\rm BiS_2}$ layers, close to the Fermi energy, is described by a two-dimensional, two-orbital 
tight-binding model, which is extracted from first principles Density Functional Theory calculations 
by using maximally localized Wannier orbitals centered at the Bismuth sites \cite{Usui2012a}. 
These Wannier states originate from the Bismuth $6p_X$ and $6p_Y$ orbitals. In reciprocal space, the 
tight-binding Hamiltonian can be written as:

\begin{eqnarray}\label{kspace}
H_0 &=& \sum_{\mathbf{k},\sigma=\uparrow \downarrow} \sum_{\alpha,\beta=X,Y} T^{\alpha\beta}
(\mathbf{ k})
p^\dagger_{\alpha \mathbf{ k} \sigma} p_{\beta \mathbf{ k} \sigma},
\end{eqnarray}
where 
\begin{eqnarray}
T^{XX} &=& 2t_x^X\left(\cos k_x +\cos k_y\right) +2t_{x \mp y}^{X} \cos  \left(k_x \pm k_y\right)  \nonumber \\ 
&+& 2t_{2x \mp y}^{X}\left[\cos  \left(2k_x \pm k_y\right) + \cos  \left(k_x \pm 2k_y\right) \right]+\epsilon_X - \mu,  \nonumber \\
T^{YY} &=& 2t_x^Y\left(\cos k_x +\cos  k_y\right) +2t_{x \pm y}^{Y} \cos  \left(k_x \mp k_y\right)  \nonumber \\
&& 2t_{2x \pm y}^{Y}\left[\cos  \left(2k_x \mp k_y\right) + \cos  \left(k_x \mp 2k_y\right) \right]+\epsilon_Y - \mu,  \nonumber \\
T^{XY} &=& T^{YX}  \nonumber \\ 
&=& 2t_x^{XY}\left(\cos k_x -\cos  k_y\right) + 4t_{2x}^{XY}\left(\cos 2k_x -\cos  2k_y\right)  \nonumber \\ 
&+& 4t_{2x+y}^{XY}\left(\cos 2k_x\cos  k_y - \cos  k_x \cos 2k_y\right).  \label{eqt12}
\end{eqnarray}
The operator $p^{\dagger}_{\alpha \mathbf{ k} \sigma}$ ($p_{\alpha \mathbf{ k} \sigma}$) in 
eq.~(\ref{kspace}) creates (annihilates)  
an electron in a Bloch state with orbital character $\alpha=X,Y$, 
with spin $\sigma=\uparrow \downarrow$, and momentum $\mathbf{ k}$. 
The values for the hopping parameters are those from Ref.~\cite{Usui2012a}, and are reproduced 
in Table 1 for convenience. Note that the choice of an upper or lower sign in the $\pm$ and $\mp$ 
in the arguments of the trigonometric functions in the equations above will determine the choice of the 
corresponding hopping parameters that also have $\pm$ and $\mp$ in their subindexes. 
It is important to note that, following Ref.~\cite{Usui2012a}, we denote the $p_X$ and $p_Y$ Wannier 
orbitals using uppercase letters ($X,Y$) and the crystallographic axes by lowercase ones ($x,y$) as they 
are rotated in relation to each other by $\pi/4$ \cite{Usui2012a}, i.e., the Wannier orbitals $p_X$ and $p_Y$ 
are oriented along the diagonals of the square lattice defined by the crystallographic axes. 

\begin{figure}
\centering
\begin{minipage}{0.45\textwidth}
\centering
\includegraphics[width =\textwidth]{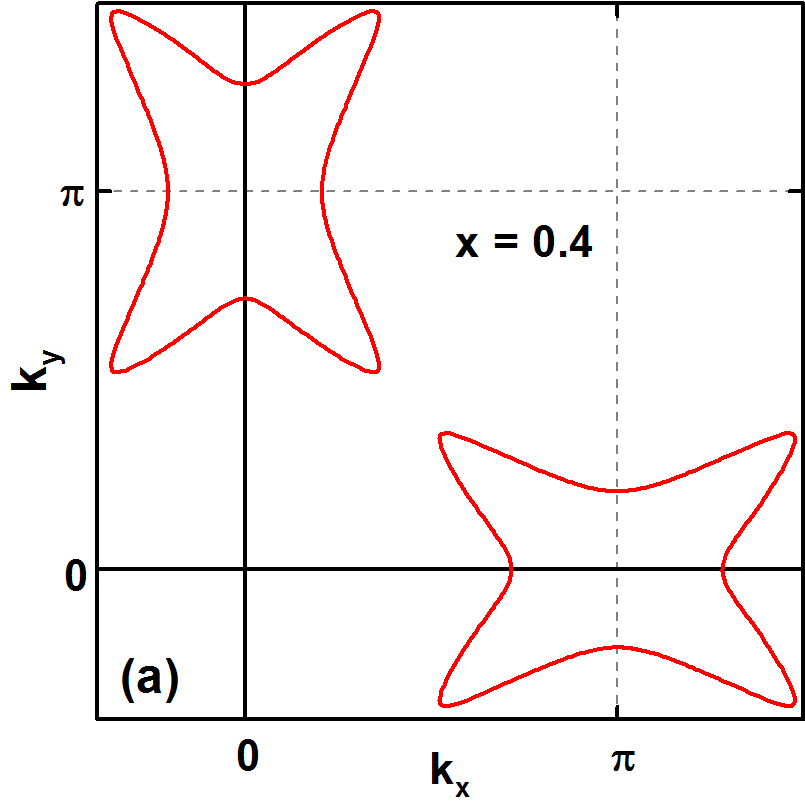}
\end{minipage}
\hfill
\begin{minipage}{0.45\textwidth}
\centering
\includegraphics[width =\textwidth]{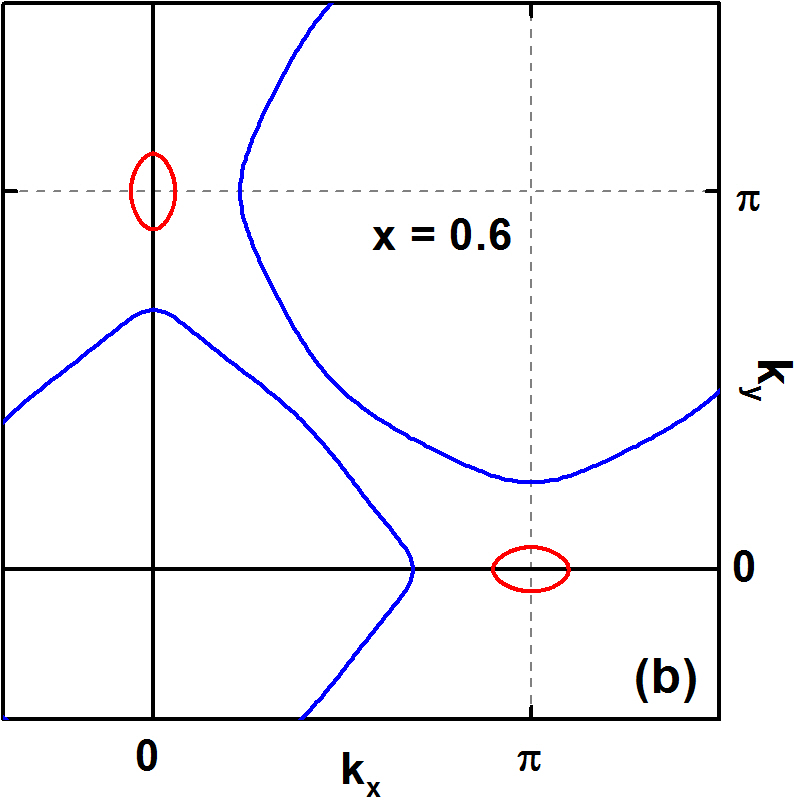}
\end{minipage}
	\caption{(Color online) Fermi surface at different electron doping values (a) $\x = 0.4$ and 
	(b) $\x = 0.6$. In panel (a), the electron pockets (red) around $[\pi,0]$ and 
	$[0,\pi]$ will keep increasing in area (as $\x$ increases) until they touch (for $\x \approx 0.45$), 
	when a Lifshitz transition occurs and the Fermi surface changes topology to 
	hole pockets (blue) centered around $[0,0]$ and $[\pi,\pi]$ [panel (b)], which will decrease in area as 
	$\x$ increases. Finally, as shown in panel (b), small electron pockets (red) 
	emerge around $[\pi,0]$ and $[0,\pi]$ (for $\x \approx 0.5$) and will increase in area as $\x$ increases. 
}
\label{figure1}
\end{figure}

The chemical potential $\mu$ varies with electron doping $\x$ and will control the filling of the bands, where $\x=0$ 
indicates that the bands are empty and $\x=1$ represents quarter-filling (i.e., 1 electron, out of a maximum of 4, per site). 
Figure \ref{figure1} shows the Fermi surface for two different values of doping, $\x=0.4$ in panel (a), 
where we see two electron pockets (red) around $[\pi,0]$ and $[0,\pi]$, which grow with $\x$. At $\x \approx 0.45$ 
they will touch and the Fermi surface will undergo a Lifshitz transition to two hole pockets centered 
around $[0,0]$ and $[\pi,\pi]$. These are shown (blue) in panel (b) for $\x=0.6$. In addition, for $\x \approx 0.5$ 
two electron pockets (red) around $[\pi,0]$ and $[0,\pi]$ will emerge and grow with $\x$, while the hole 
pockets decrease. 

\subsection{Interacting Hamiltonian}\label{inth}

For our model of  ${\rm BiS_2}$-based superconductors, we will assume that attractive interactions  mediate 
different types of intra- and inter-orbital pairings \cite{Suhl1959}. The relevant 
orbitals, as discussed above, are the  $p_{X}$ and $p_{Y}$ Wannier orbitals of the Bismuth atoms 
in a ${\rm BiS}$ plane. 
The total Hamiltonian of the system can be written as 
\begin{equation}
H=H_0+H_{I}, 
\end{equation}
where $H_{0}$ is given by eqs.~(\ref{kspace}) and (\ref{eqt12}) above and the  
interacting part of the Hamiltonian can be written as  a 
sum of intra- and inter-orbital components,  
\begin{eqnarray*}\label{HEP}
H_{I} & = &H_{\rm intra}+H_{\rm inter},
\end{eqnarray*}
where
\begin{eqnarray}\label{Hintraband}
H_{\rm intra}& = & -\frac{1}{N}\sum_{k,k^{\prime},\alpha}\left( V_{\alpha \alpha \alpha \alpha}~
p_{\alpha k^{\prime} \uparrow}^{\dagger}p_{\alpha \bar{k}^{\prime} \downarrow}^{\dagger} 
p_{\alpha \bar{k} \downarrow}p_{\alpha k \uparrow}\right) ,  
\end{eqnarray}

\begin{eqnarray}\label{Hinterband}
H_{\rm inter} & = & -\frac{1}{N} \sum_{k,k^{\prime},\alpha \neq \beta} \Big(V_{\alpha \alpha \beta \beta}~
p_{\alpha k^{\prime} \uparrow}^{\dagger} p_{\alpha \bar{k}^{\prime} \downarrow}^{\dagger}
p_{\beta \bar{k} \downarrow}p_{\beta k \uparrow}  \nonumber \\
	& + &V_{\alpha \beta \beta \alpha}~p_{\alpha k^{\prime} \uparrow}^{\dagger} p_{\beta \bar{k}^{\prime} \downarrow}^{\dagger}
p_{\beta \bar{k} \downarrow}p_{\alpha k \uparrow} 
+ V_{\alpha \beta \alpha \beta}~p_{\alpha k^{\prime} \uparrow}^{\dagger} p_{\beta \bar{k}^{\prime} \downarrow}^{\dagger}
p_{\alpha \bar{k} \downarrow}p_{\beta k \uparrow}\Big), 
\end{eqnarray}
where  $\alpha,\beta=X,Y$, $\bar{k} = -k$, and $N=L^2$ is the number of sites in a $L \times L$ square lattice. 
To be accurate, as we chose the \emph{orbital} states $p_{X,Y \mathbf{k} \sigma}$ to write the pair operators, we 
will use the terminology intra- and inter-orbital to refer to the associated pairing, in opposition to intra and interband. 
The main reason for using the $p_{X,Y \mathbf{k} \sigma}$-orbital states to write the pair operators is that the 
actual bands [obtained by diagonalizing $H_{0}$] show weak $X-Y$ hybridization, because of the small 
value of $t_x^{XY}=0.107$ in comparison to $t_x^{X,Y}=0.880$ (see Table 1). In addition, in systems where many-body terms 
originating from intra-site interactions may influence superconductivity (see, for example, Ref.~\cite{Graser2009}), 
which could be the case for ${\rm BiS_2}$ compounds \cite{Hirsch2015}, it is advantageous to analyze pairing in the orbital basis.

In the equations above, all the coupling terms $V$ are positive, therefore all pairing interactions considered are attractive. 
In addition, experimental findings for ${\rm BiS_2}$ compounds \cite{Yazici2015}, up to now, support s-wave SC, 
therefore, we take all the coupling terms as being $\mathbf{ k}$-independent. 
Given that the origin of the pairing interaction in ${\rm BiS_2}$ compounds has not been settled yet \cite{Hirsch2015}, 
those are the only general assumptions we will make. In the next two sections we will use symmetry 
arguments to decrease the number of $V$ terms in eqs.~(\ref{Hintraband}) and (\ref{Hinterband}) when applied to 
${\rm BiS_2}$ compounds. 

The inter-orbital terms in eq.~(\ref{Hinterband}) 
may be listed through the associated couplings as 
${\rm V_{XXYY}}$ (${\rm V_{YYXX}}$), where an ${\rm YY}$ (${\rm XX}$) pair is scattered into an ${\rm XX}$ 
(${\rm YY}$) pair; ${\rm V_{XYYX}}$ (${\rm V_{YXXY}}$), where an ${\rm YX}$ (${\rm XY}$) pair is scattered into an 
${\rm XY}$ (${\rm YX}$) pair; and ${\rm V_{XYXY}}$ (${\rm V_{YXYX}}$), where an ${\rm XY}$ (${\rm YX}$) pair is 
scattered into an ${\rm XY}$ (${\rm YX}$) pair. 
We will see in what follows that, by treating $H_{\rm intra}$ and $H_{\rm inter}$ 
at the mean-field level, and applying symmetries present in the ${\rm BiS}$ planes, will allow us 
to reduce these couplings to just two, which we will denote as ${\rm V_{XX}}$ and ${\rm V_{XY}}$. 

\begin{figure}
\centering
\includegraphics[width =3.8in]{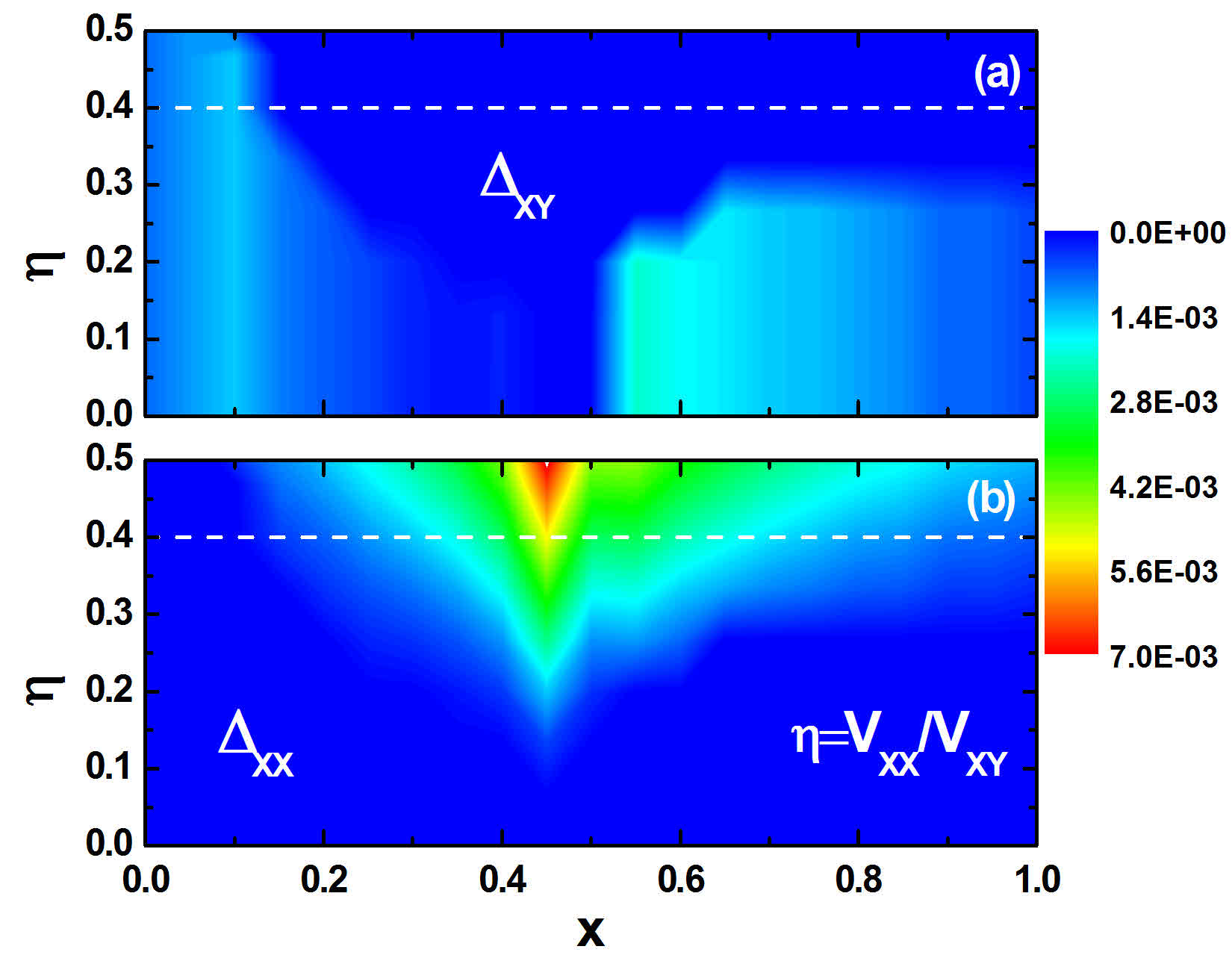}
\caption{(Color online) ${\rm \Delta_{XY}}$ (a) and ${\rm \Delta_{XX}}$ (b) phase diagram 
for ${\rm \eta=V_{XX}/V_{XY}}$ {\em vs} doping $\x$ (gap values in eV). These results 
	were obtained through the self-consistent solution of eqs.~\ref{D2new} and \ref{D3new} 
	and the parameters used were $V_{XY}=0.19$~eV and $\omega_D=10$~meV, which was used
	as a cutoff for energy integrals in solving the gap equations. 
}
\label{figure2}
\end{figure}

\subsection{Mean-field theory and gap equations}\label{gapeq}

The interacting Hamiltonian in eqs.~(\ref{Hintraband}) and (\ref{Hinterband}) will be solved 
at the mean-field level, through the usual approximation 
$ABCD \approx \langle AB \rangle CD + AB \langle CD \rangle - \langle AB \rangle \langle CD \rangle$: 
\begin{eqnarray}\label{HEP}
	H_{I} \approx H_{MF} &=& H_{1}+H_{2}+C, 
\end{eqnarray}
where 
\begin{eqnarray}\label{h1}
	H_{1}&=& -\sum_{k}(\Delta_{1}~p^{\dagger}_{Xk\uparrow}p^{\dagger}_{X\bar{k}\downarrow}
+\Delta_{2}~p^{\dagger}_{Yk\uparrow}p^{\dagger}_{Y\bar{k}\downarrow}+h.c.) 
\end{eqnarray}
and
\begin{eqnarray}\label{h2}
	H_{2}&=&  -\sum_{k}( \Delta_{3}~p^{\dagger}_{Xk\uparrow}p^{\dagger}_{Y\bar{k}\downarrow}+
\Delta_{4}~p^{\dagger}_{Yk\uparrow}p^{\dagger}_{X\bar{k}\downarrow}+h.c.), 
\end{eqnarray}
where $C$ is a constant and the $\Delta_{i}$, which are order parameters of the 
superconducting phases \cite{note2}, are given by  
\begin{eqnarray}
\Delta_{1}&=& \frac{1}{N}\sum_{k^{\prime}}V_{XXXX}\langle p_{X\bar{k}^{\prime}\downarrow}p_{Xk^{\prime}\uparrow}\rangle +
V_{XXYY}\langle p_{Y\bar{k}^{\prime}\downarrow}p_{Yk^{\prime}\uparrow}\rangle,   \label{d1} \\
\Delta_{2}&=& \frac{1}{N}\sum_{k^{\prime}}V_{YYYY}\langle p_{Y\bar{k}^{\prime}\downarrow}p_{Yk^{\prime}\uparrow}\rangle +
V_{YYXX}\langle p_{X\bar{k}^{\prime}\downarrow}p_{Xk^{\prime}\uparrow}\rangle,  \label{d2} \\
\Delta_{3}&=& \frac{1}{N}\sum_{k^{\prime}}V_{XYYX}\langle p_{Y\bar{k}^{\prime}\downarrow}p_{Xk^{\prime}\uparrow}\rangle +
V_{XYXY}\langle p_{X\bar{k}^{\prime}\downarrow}p_{Yk^{\prime}\uparrow}\rangle,  \label{d3} \\
\Delta_{4}&=& \frac{1}{N}\sum_{k^{\prime}}V_{YXXY}\langle p_{X\bar{k}^{\prime}\downarrow}p_{Yk^{\prime}\uparrow}\rangle +
V_{YXYX}\langle p_{Y\bar{k}^{\prime}\downarrow}p_{Xk^{\prime}\uparrow}\rangle. \label{d4}
\end{eqnarray}

\subsection{Applying symmetries}\label{symm}

We start our analysis from the fact that ${\rm V_{XXYY}=V_{YYXX}}$ 
and ${\rm V_{XYYX}=V_{YXXY}}$. 
Now, to simplify eqs.~(\ref{d1}) to (\ref{d4}), we will apply some symmetry properties 
related to the Bismuth $p_X$ and $p_Y$ orbitals, which lead to relations between 
the remaining couplings ${\rm V}$ and between the expectation values in those equations.
Given that both orbitals have the same energy and 
are related by a $C_4$ rotation \cite{Usui2012a}, we expect that ${\rm V_{XXXX}= V_{YYYY}}$ 
and $ \langle p_{X\bar{k}^{\prime}\downarrow}p_{Xk^{\prime}\uparrow}\rangle = 
\langle p_{Y\bar{k}^{\prime}\downarrow}p_{Yk^{\prime}\uparrow}\rangle$. Therefore, $\Delta_1 = \Delta_2$
(which we now denote as ${\rm \Delta_{XX}}$), and, if we define ${\rm V_{XX} \equiv V_{XXXX} + V_{XXYY}}$, we obtain
\begin{eqnarray} \label{D2new}
\Delta_{XX}&=&-\frac{V_{XX}}{N}\sum_{k^{\prime}}\langle p_{X\bar{k}^{\prime}\downarrow}
p_{Xk^{\prime}\uparrow} \rangle.
\end{eqnarray}
Note that the above equation replaces eqs. (\ref{d1}) and (\ref{d2}).
The same symmetry arguments lead to ${\rm V_{XYXY}=V_{YXYX}}$ 
and $\langle p_{Y\bar{k}^{\prime}\downarrow}p_{Xk^{\prime}\uparrow}\rangle = 
\langle p_{X\bar{k}^{\prime}\downarrow}p_{Yk^{\prime}\uparrow}\rangle$, which result 
in $\Delta_3 = \Delta_4$ (which we now denote as ${\rm \Delta_{XY}}$), 
and, if we define ${\rm V_{XY} \equiv V_{XYYX} + V_{XYXY}}$, we obtain
\begin{eqnarray} \label{D3new}
\Delta_{XY}&=&-\frac{V_{XY}}{N}\sum_{k^{\prime}}\langle p_{Y\bar{k}^{\prime}\downarrow}
	p_{Xk^{\prime}\uparrow} \rangle.
\end{eqnarray}
Note that the above equation replaces eqs. (\ref{d3}) and (\ref{d4}).

The simplified gap equations (\ref{D2new}) and (\ref{D3new}) determine the system 
of self-consistent equations to be solved,  
where the effective interactions ${\rm V_{XX}}$ and ${\rm V_{XY}}$ are parameters that control 
pairing of same-orbital electrons (${\rm XX}$ or ${\rm YY}$), or dissimilar electrons (${\rm XY}$ or ${\rm YX}$), respectively. 
We stress that the order parameter ${\rm \Delta_{XX}}$ involves both intra-orbital (${\rm XX \leftrightarrow XX}$ 
and ${\rm YY \leftrightarrow YY}$) as well as inter-orbital (${\rm XX \leftrightarrow YY}$) processes, while the order 
parameter ${\rm \Delta_{XY}}$ involves only inter-orbital processes (${\rm XY \leftrightarrow YX}$, ${\rm XY \leftrightarrow XY}$, 
and ${\rm YX \leftrightarrow YX}$). 

Figure \ref{figure2} shows the results obtained by self-consistently solving eqs.~(\ref{D2new}) and (\ref{D3new}) for 
both ${\rm \Delta_{XY}}$, in panel (a), and ${\rm \Delta_{XX}}$, in panel (b). 
The phase diagram presents a color map plot of both gap functions in the $\eta$ {\em vs} $\x$ plane, 
where ${\rm \eta=V_{XX}/V_{XY}}$ measures the ratio between the 
couplings (${\rm V_{XY} = 0.19~eV}$ was kept constant while ${\rm V_{XX}}$ varied). 
The color scale is the same for both panels and it is given in eV units. 
Details of the self-consistent numerical solution of the gap equations can be found in Ref.~[\cite{Griffith2016}]. 

\begin{figure}
\centering
\includegraphics[width =3.8in]{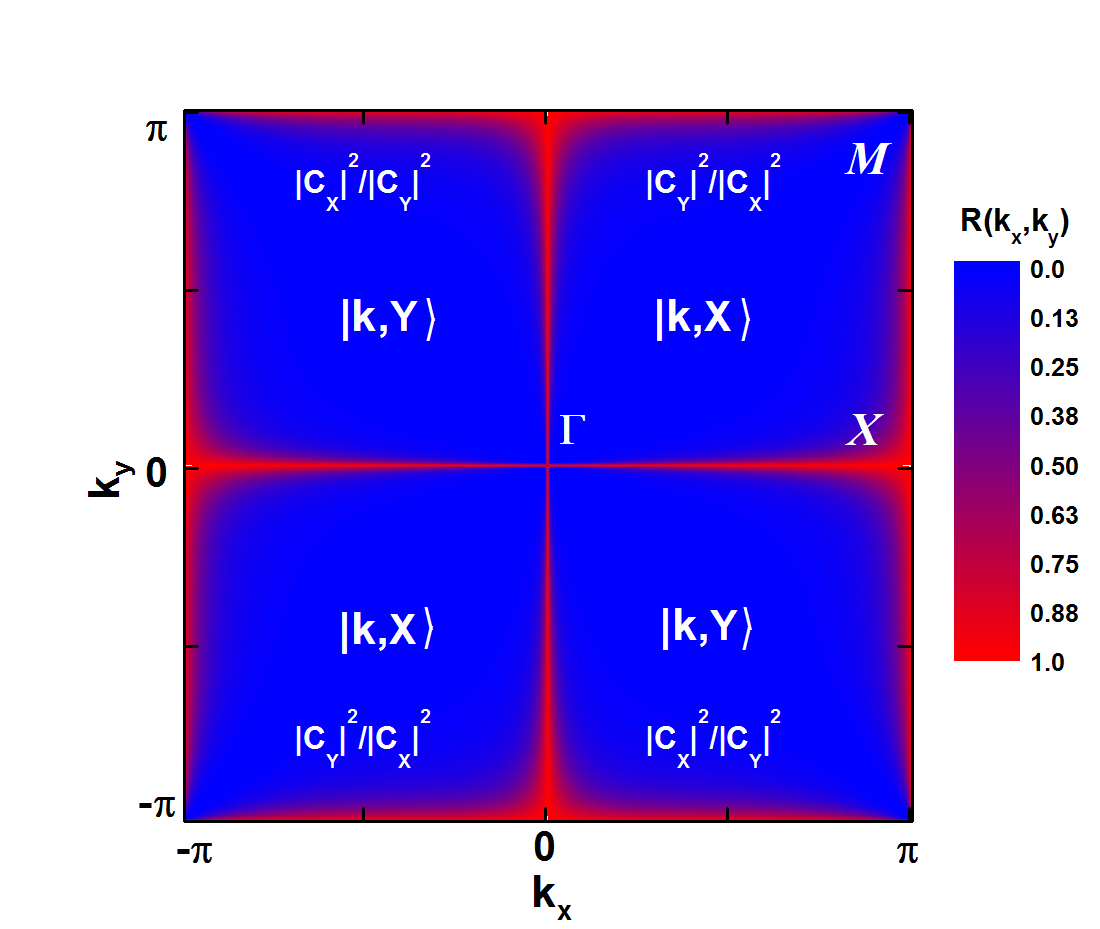}
\caption{(Color online) Variation along the BZ of the orbital-mixing ratio 
	$R(k_x,k_y)=|C_{X,Y}|^2/|C_{Y,X}|^2$ for the lower energy band, showing that 
	a well defined orbital character (either $|k,X\rangle$ or $|k,Y\rangle$, for $R(k_x,k_y)\approx 0$) 
	alternates from one quadrant to the next, with the orbital mixed character ($R(k_x,k_y)\approx 1$) 
	concentrated along the crystallographic axes. Note that, as indicated in the figure, the choice of 
	the $X$ and $Y$ subindexes in the definition of $R(k_x,k_y)$ for each quadrant is such that 
	$0 \leq R(k_x,k_y) \leq 1$ for the whole BZ. The results for the higher energy band 
	are identical to the ones shown here, but with $X$ and $Y$ swapped. 
}
\label{figure3}
\end{figure}

\section{Results and Discussion}

\subsection{Orbital-mixing and multiband superconductivity}\label{ratiosec}

The superconducting state emerges from an instability of the Fermi sea (metallic normal state) 
to an attractive effective interaction. This interaction forms Cooper pairs that scatter against 
each other, always conserving total momentum and individual spin, while staying 
in a shell around the Fermi surface. It is then expected that many properties of the superconducting state, 
like the gap function and, in a multiband system, the possibility of the existence of 
different types of Cooper pairs, will be directly associated to the properties of the Fermi surface 
in the normal state. Thus, in a system like ${\rm BiS_2}$, whose Fermi surface varies widely with 
electron doping, even showing a Lifshitz transition, as illustrated in Fig.~\ref{figure1}, one 
would expect that the superconducting gap function should 
also show a marked variation with electron doping. Indeed, the gap function results in Fig.~\ref{figure2} 
clearly confirm this expectation by showing very marked variations when the system 
goes through the Lifshitz transition (for $\x \approx 0.45$). However, as described in this section, our results, 
when analyzed more carefully, also show that there is a more 
subtle aspect relating multiband SC with the nature of the band states at the Fermi surface. 
This aspect, once properly quantified, can be directly linked to the very 
particular electronic structure of ${\rm BiS_2}$ compounds. 
As will be illustrated below, the orbital-mixing character of a band state
changes from point to point in the BZ of ${\rm BiS_2}$, varying continuously from pure-${\rm X}$ to pure-${\rm Y}$ 
(and back, passing by completely-${\rm XY}$-mixed) in accordance to symmetry requirements. 
As a consequence, the degree of mixing of the ${\rm X}$- and ${\rm Y}$-orbital at the Fermi surface, for a particular 
electron doping, may change between different regions of the Fermi surface. This is not surprising in itself. 
What is interesting in the case of ${\rm BiS_2}$ is that the systematic variation of 
the orbital-mixing along the ${\rm BiS_2}$ Fermi surface can be 
semi-quantitatively connected to the ${\rm \Delta_{XY}}$ and ${\rm \Delta_{XX}}$ results in Fig.~\ref{figure2}. 

Therefore, we will use the idea of orbital-mixing, as defined below,
as well as the way the Fermi surface changes with doping, to explain the main 
structures seen in the superconducting gap functions shown in Fig.~\ref{figure2}, as, 
for example, the position of the maxima and minima of ${\rm \Delta_{XY}}$ and 
${\rm \Delta_{XX}}$ as a function of electron doping $\x$. 
Note that we assume a rigid band situation (i.e., doping does not 
change the band structure); ARPES results \cite{Yazici2015} 
have shown that this is a good approximation for ${\rm BiS_2}$ compounds.  

\begin{figure}
\centering
\begin{minipage}{0.45\textwidth}
\centering
\includegraphics[width =\textwidth]{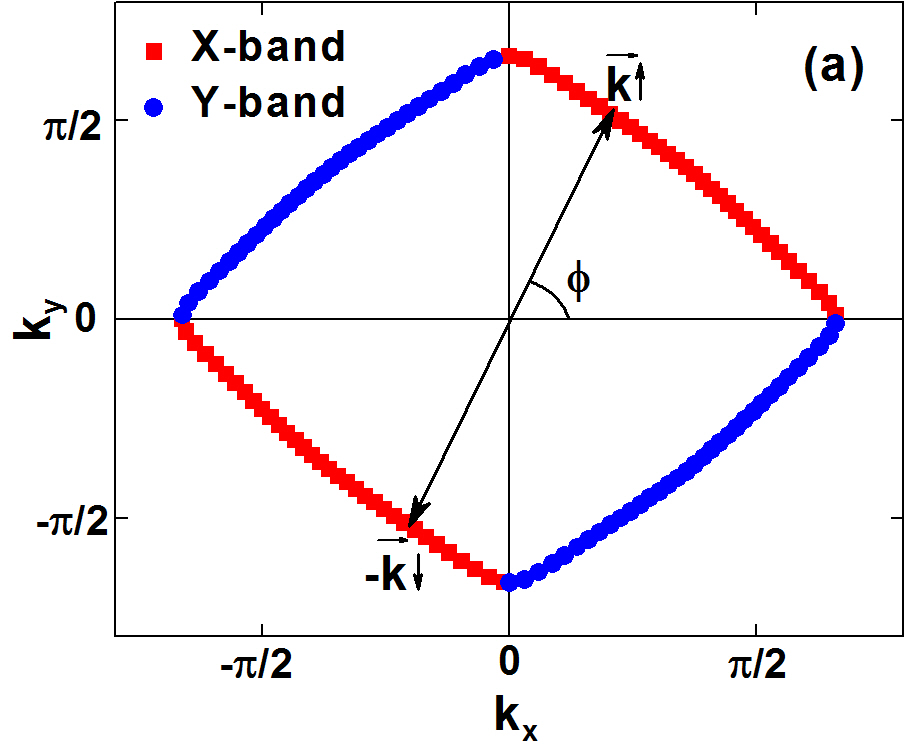}
\end{minipage}
\hfill
\begin{minipage}{0.45\textwidth}
\centering
\includegraphics[width =\textwidth]{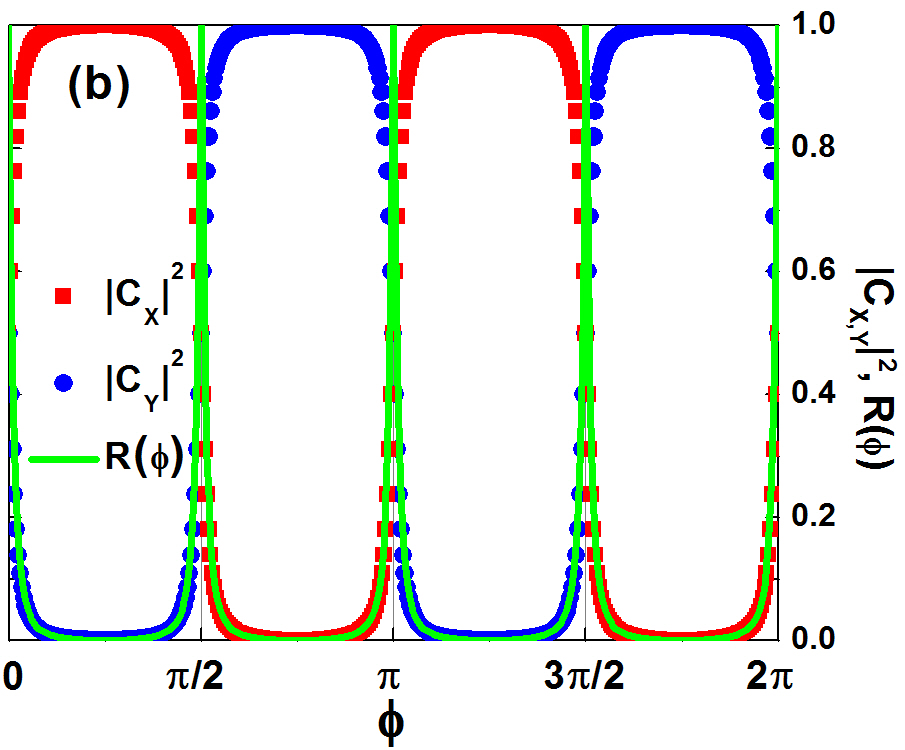}
\end{minipage}
\caption{(Color online) (a) Example of a hole pocket centered around $[0,0]$ with almost zero-mixing ($\x=0.75$): 
the well defined ${\rm X}$-orbital [(red) squares] or ${\rm Y}$-orbital [(blue) circles] character of the Fermi surface 
states changes from one quadrant to the next. This well defined character is determined 
by $|C_{X,Y}|^2 \approx 1.0$, where a generic band state at the Fermi surface 
is written as $|\vec{k} \rangle = C_X|p_{k,X} \rangle  + C_Y|p_{k,Y} \rangle$. 
(b) $|C_{X}|^2$ [(red) squares]  and $|C_{Y}|^2$ [(blue) circles] values around the Fermi surface 
($0 \leq \phi \leq 2\pi$) for the hole pocket in panel (a). The solid (green) curve shows 
the corresponding results for $R(\phi)$, the orbital-mixing ratio, which vanishes for most of the Fermi surface, 
aside from very small regions around multiples of $\pi/2$ where symmetry imposes an ${\rm X \leftrightarrow Y}$ 
swap of the orbital character of the band states. 
In panel (a), a generic  Cooper pair will be of ${\rm XX}$- (as depicted) 
or ${\rm YY}$-type, no ${\rm XY}$-type being possible.
}
\label{figure4}
\end{figure}

Consider a band state, at a generic $\mathbf{k}=[k_x,k_y]$ in the first BZ, 
written as $|\mathbf{k} \rangle = C_X(k_x,k_y)|p_{X,k} \rangle  
+ C_Y(k_x,k_y)|p_{Y,k} \rangle$, where $|p_{\alpha,k} \rangle = p^{\dagger}_{\alpha \mathbf{k}} |vac \rangle$, 
for $\alpha = X$, $Y$ (where the spin index was omitted for the sake of brevity). 
To quantify the degree of orbital-mixing, we define $R(k_x,k_y) = |C_{\alpha}(k_x,k_y)|^2/|C_{\beta}(k_x,k_y)|^2$, where
$\alpha$ and $\beta$ take values $X$ or $Y$ such that $0.0 \leq R(k_x,k_y) \leq 1.0$ for all $[k_x,k_y]$. 
Thus, we refer to $R(k_x,k_y)$ as the \emph{orbital-mixing ratio} between the ${\rm X}$-orbital and ${\rm Y}$-orbital 
for each point of the BZ, where $R(k_x,k_y)=1.0$ indicates maximum orbital mixing, where the 
band state does not have a well defined ${\rm X}$- or ${\rm Y}$-orbital character, being an equal mix of both; 
and $R(k_x,k_y)=0.0$ indicates no orbital mixing at all, 
i.e., the band state has a well defined (either ${\rm X}$- or ${\rm Y}$-) orbital character. We will refer to the 
former as a \emph{orbital-mixed} band state and to the latter as a \emph{zero-mixing} band state. 

Figure \ref{figure3} shows a color-map plot of $R(k_x,k_y)$ for the lower energy band in the first BZ. 
As indicated by the labels, a well defined orbital character can be associated to the band states 
(denoted as either $|k,X \rangle$, when $|C_{X}(k_x,k_y)| \approx 1$, or $|k,Y \rangle$, when $|C_{Y}(k_x,k_y)| \approx 1$) 
in a wide range around the $\Gamma M$ symmetry lines (where $R(k_x,k_y) \approx 0$) 
in each quadrant (alternating from $|k,X \rangle$ to $|k,Y \rangle$ from one quadrant to the next), 
while the band states in a narrow region around the $\Gamma X$ and $XM$ symmetry lines are orbital-mixed ($R(k_x,k_y) \approx 1$). 
Results for the higher energy band are identical, but for the swapping of $X$ and $Y$. 
Based on these results, parts of the Fermi surface that are formed by large hole pockets 
around the $\Gamma$ and \emph{M} points in the BZ [see Fig.~\ref{figure1}(b)] contain mostly 
band states with zero-mixing, while parts of the Fermi surface formed by smaller electron pockets 
around the \emph{X} points in the BZ will contain band states with a larger degree of orbital mixing. 
In what follows, we will analyze the orbital-mixing ratio $R(k_x,k_y)$ over Fermi surface pockets, i.e., 
we will be interested on the variation of $R(\phi)$ as we move around the pocket's edge, where we have parametrized 
$\vec{k}_F$ as $[k_F,\phi]$, where $\phi$ is the polar angle measured around the pocket's center. 

\begin{figure}
\centering
\begin{minipage}{0.45\textwidth}
\centering
\includegraphics[width =\textwidth]{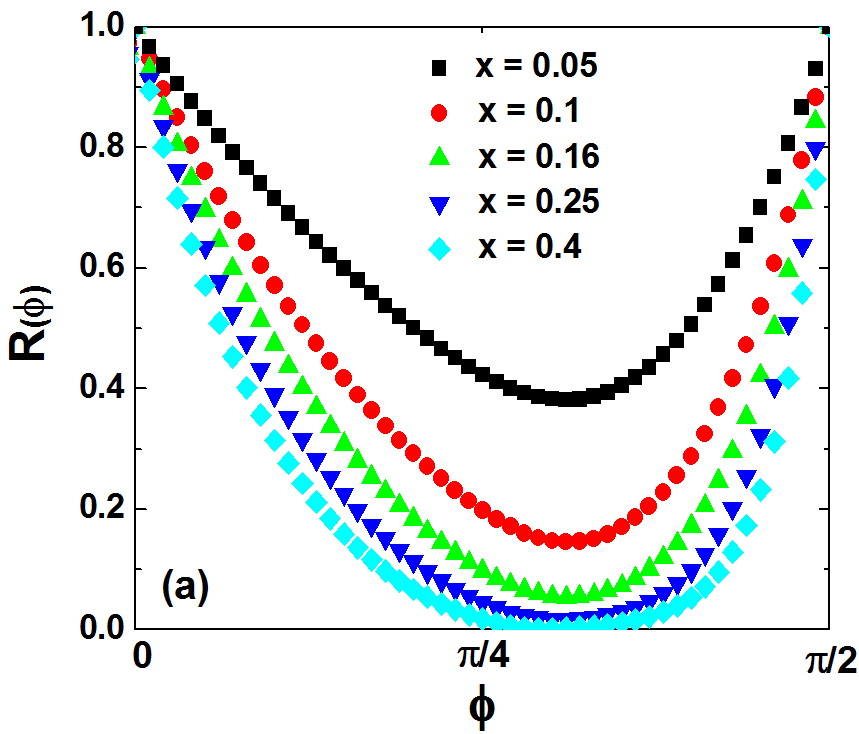}
\end{minipage}
\hfill
\begin{minipage}{0.42\textwidth}
\centering
\includegraphics[width =\textwidth]{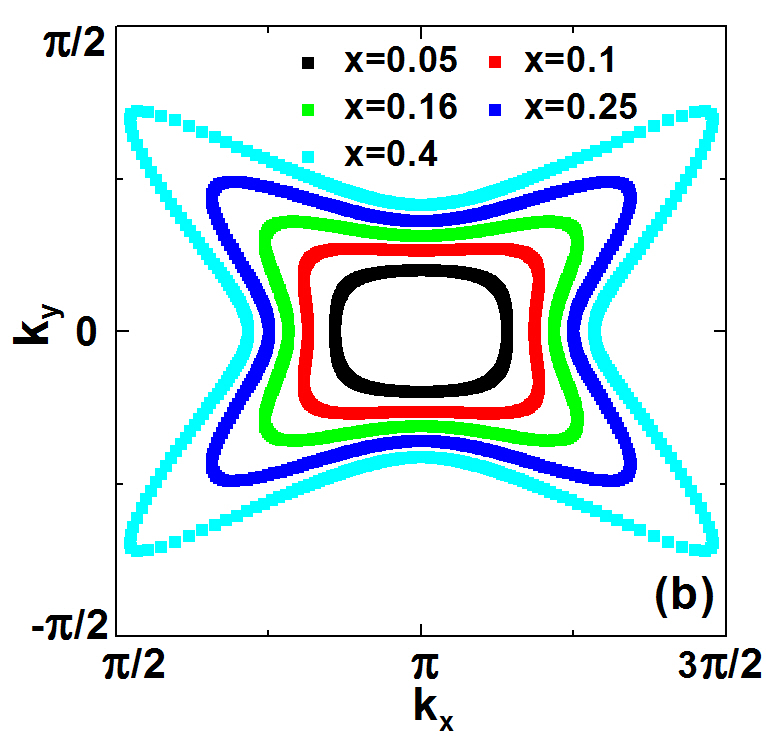}
\end{minipage}
\caption{(Color online) (a) $R(\phi)$ (see text for definition) in the first quadrant of 
the $[\pi,0]$ electron pocket for $\x=0.05$ [(black) squares], $\x=0.1$ [(red) circles], $\x=0.16$ 
[(green) up triangles], $\x=0.25$ [(blue) down triangles], and $\x=0.4$ [(cyan) diamonds]. 
(b) $[\pi,0]$ electron pockets for the same values of $\x$ as in panel (a). 
}
\label{figure5}
\end{figure}

To clarify the connection between orbital-mixing and Cooper pair formation in a multiband system and therefore 
develop an intuitive picture of the relation between the ${\rm \Delta_{XY,XX}}$ superconducting order parameters 
and orbital-mixing, we show in Fig.~\ref{figure4}(a) the zero-mixing case for a hole pocket around $[0,0]$ 
obtained for $\x=0.75$, with the corresponding values for $|C_{X}|^2$ [(red) squares], 
$|C_{Y}|^2$ [(blue) circles], and $R(\phi)$ [(green) solid curve] shown in Fig.~\ref{figure4}(b), 
where we see that $R(\phi) \approx 0$ for the whole pocket
with the exception of small regions around $\phi$ values that are multiples of $\pi/2$, 
where symmetry imposes an $X \leftrightarrow Y$ swap in the $|C_{X,Y}|^2$ coefficients \cite{Usui2012a}. 
Therefore, in a zero-mixing pocket, if states at the Fermi surface have very well defined 
orbital character in one specific quadrant 
[${\rm X}$-orbital, for example, in the first quadrant, as shown in Fig.~\ref{figure4}(a)], they will have the opposite 
character in the next quadrant [${\rm Y}$-orbital in the second quadrant in Fig.~\ref{figure4}(a)], and so on. 
In that case, Cooper pairs will be formed by ${\rm X}$-orbital electrons only [${\rm XX}$ pairs, like the one depicted 
in panel (a)] or ${\rm Y}$-orbital electrons only (${\rm YY}$ pairs). Exchange of 
phonons that scatter electrons between opposing quadrants will lead to ${\rm XX \leftrightarrow XX}$ and 
${\rm YY \leftrightarrow YY}$ pair scattering, while phonons that scatter electrons between
adjacent quadrants will lead to ${\rm XX \leftrightarrow YY}$ pair scattering. 
Therefore, zero-mixing pockets are associated to the SC order parameter ${\rm \Delta_{XX}}$. 
On the other hand, for electron pockets around the \emph{X} points in the BZ, where orbital-mixing dominates
i.e., $|C_{X}|^2 \approx |C_{Y}|^2 \approx 1/2$, formation of 
$p^{\dagger}_{X,k, \uparrow} p^{\dagger}_{Y,-k, \downarrow}$ pairs becomes possible, 
therefore, the order parameter ${\rm \Delta_{XY}}$ is connected to orbital-mixed pockets. 

\begin{figure}
\centering
\begin{minipage}{0.45\textwidth}
\centering
\includegraphics[width =\textwidth]{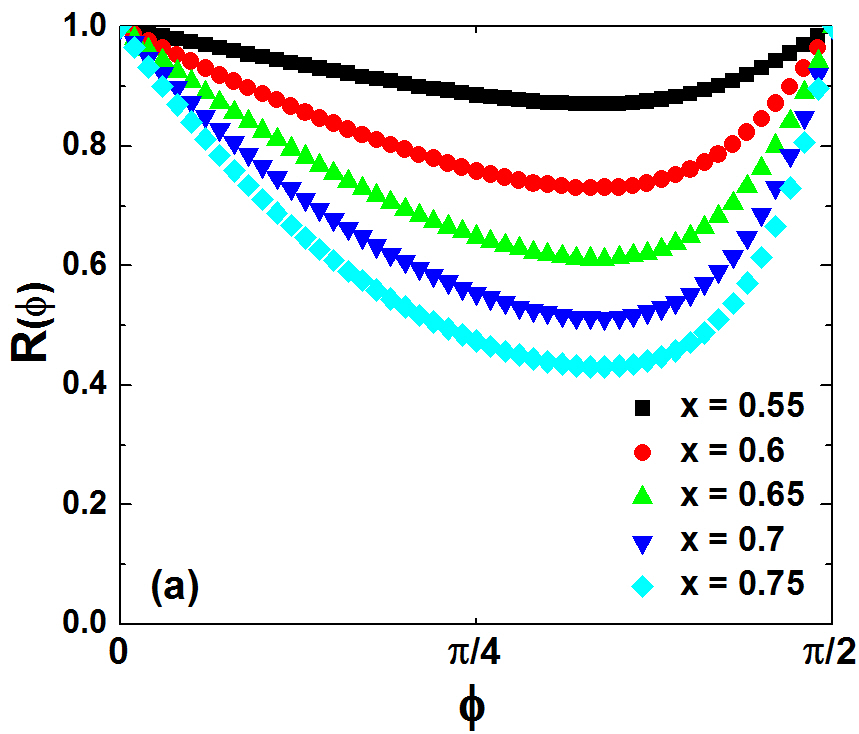}
\end{minipage}
\hfill
\begin{minipage}{0.4\textwidth}
\centering
\includegraphics[width =\textwidth]{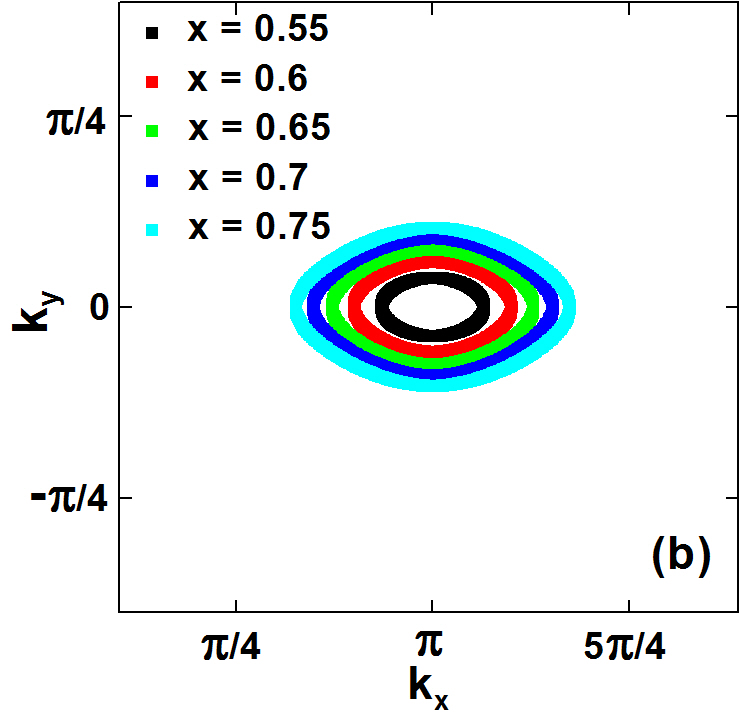}
\end{minipage}
\caption{(Color online) Same as in Fig.~\ref{figure5}, but now for 
$\x=0.55$ [(black) squares], $\x=0.6$ [(red) circles], $\x=0.16$ 
[(green) up triangles], $\x=0.7$ [(blue) down triangles], and $\x=0.75$ [(cyan) diamonds]. 
}
\label{figure6}
\end{figure}

\begin{figure}
\centering
\begin{minipage}{0.31\textwidth}
\centering
\includegraphics[width =\textwidth]{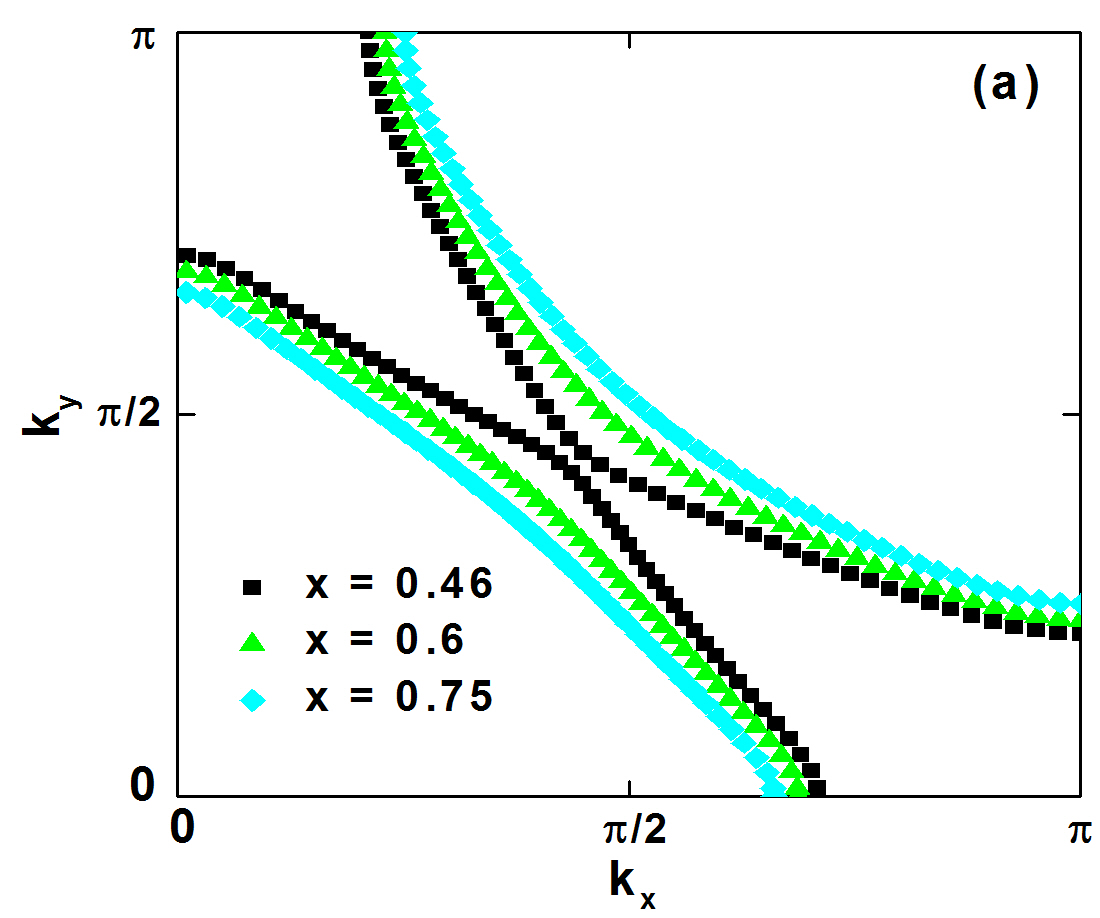}
\end{minipage}
\hfill
\centering
\begin{minipage}{0.31\textwidth}
\centering
\includegraphics[width =\textwidth]{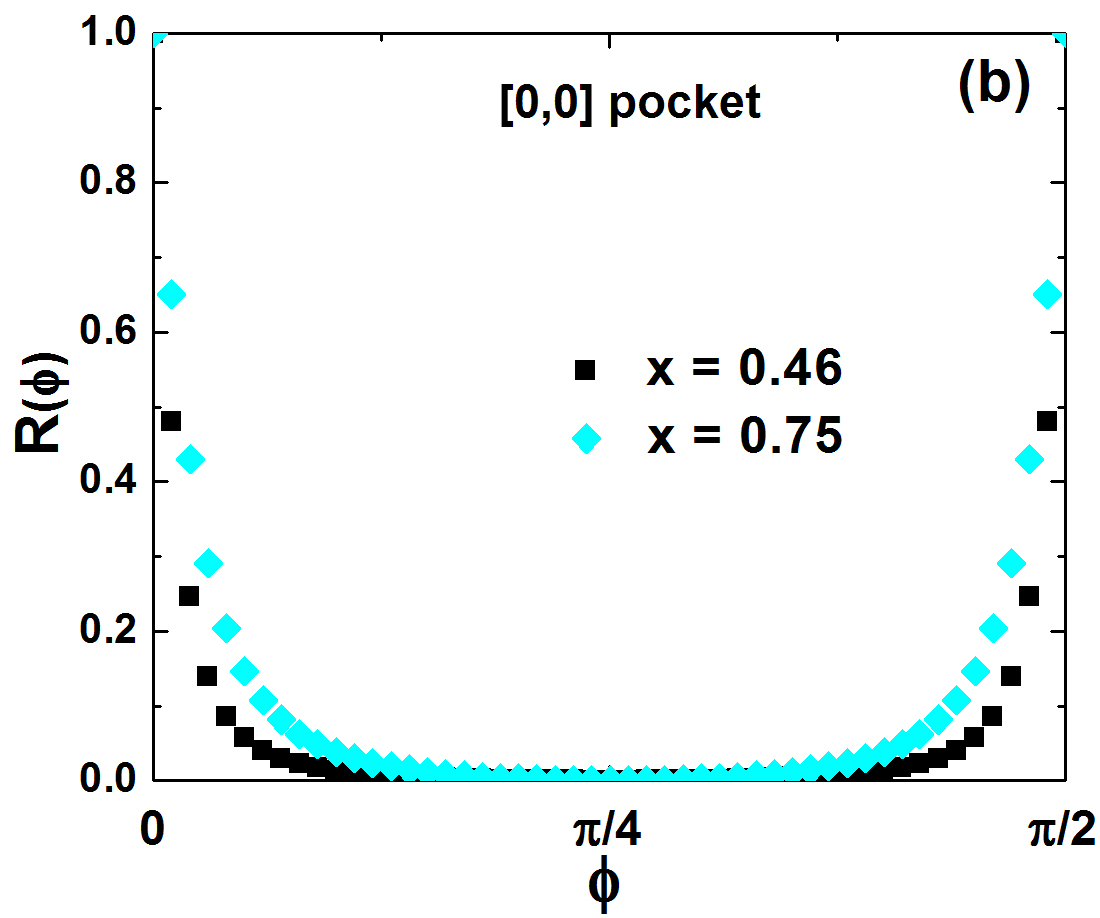}
\end{minipage}
\hfill
\begin{minipage}{0.31\textwidth}
\centering
\includegraphics[width =\textwidth]{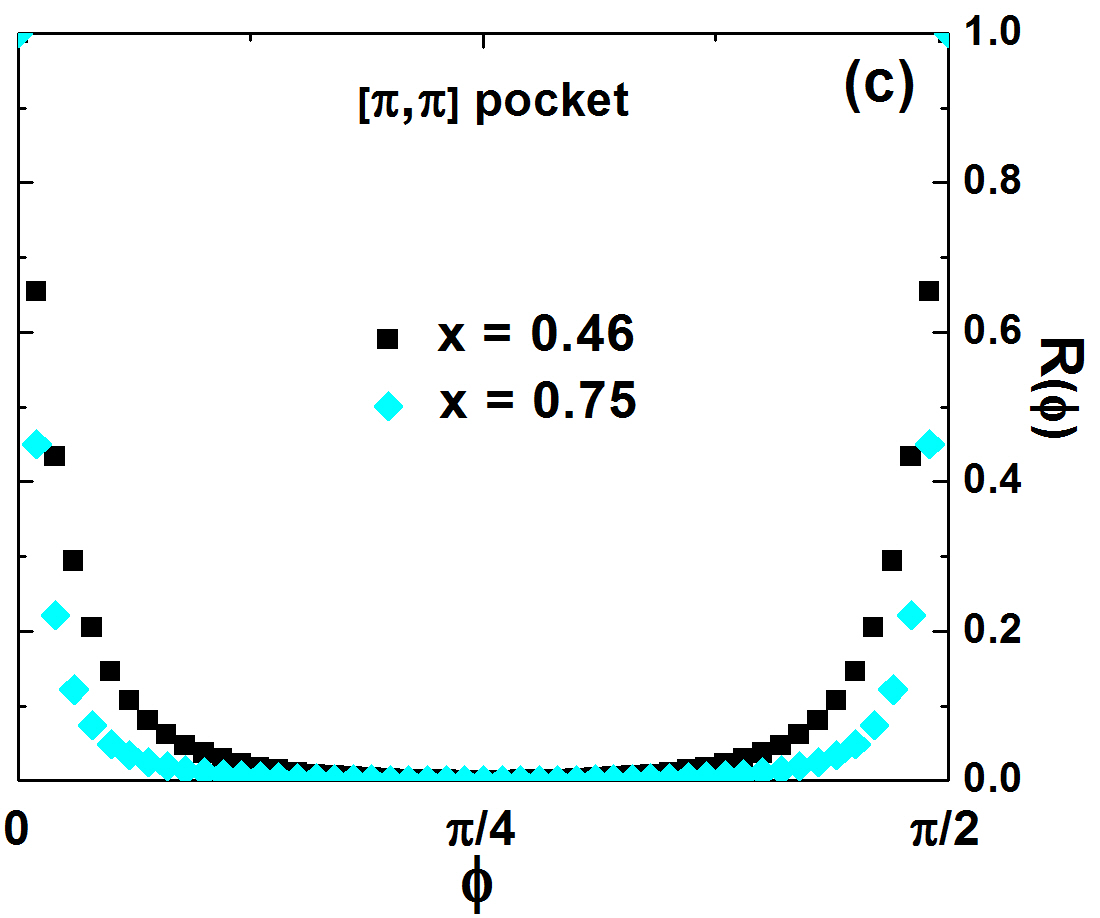}
\end{minipage}
\caption{(Color online) (a) Doping variation of the $[0,0]$ and $[\pi,\pi]$ hole pockets 
for $\x=0.46$ [(black) squares], $\x=0.6$ [(green) up triangles], and $\x=0.75$ 
[(cyan) diamonds] (b) $R(\phi)$ in the first quadrant of 
	the $[0,0]$ hole pocket [in panel (a)] for $\x=0.46$ [(black) squares] and $\x=0.75$ [(cyan) diamonds]. 
	(c) Same as in panel (b), but now for the $[\pi,\pi]$ hole pocket in panel (a). 
}
\label{figure7}
\end{figure}

It is important to realize at this point that the results shown in Fig.~\ref{figure3}, for $R(k_x, k_y)$, are obviously 
independent of the electron doping $\x$. However, since we are interested in what 
happens at the Fermi surface, as the pockets continuously expand or 
contract as $\x$ varies (see Fig.~\ref{figure1}), causing their edges to sweep through the BZ, 
$R(\phi)$ at the edge of each pocket will change substantially with electron doping for 
electron pockets centered around \emph{X} points in the BZ 
(see Figs.~\ref{figure5} and \ref{figure6}), while it will change very little for hole pockets centered 
around the $\Gamma$ and \emph{M} points in the BZ [see Fig.~\ref{figure7}(a)]. Therefore, in what follows, when we 
refer to changes in $R(\phi)$ with electron doping, that is what is meant. 

\subsection{Orbital-mixing and multiband SC below the Lifshitz transition}\label{low}

Now, using $R(\phi)$ as a measure of orbital-mixing, we will qualitatively connect 
the structures seen in the gap function results in Fig.~\ref{figure2} to the way $R(\phi)$ 
varies with doping. Let us start at $\x$ values below the Lifshitz transition, which
occurs for $\x \approx 0.45$, where the Fermi surface changes from
electron pockets centered around $[\pi,0]$ and $[0,\pi]$ to hole pockets 
centered around $[0,0]$ and $[\pi,\pi]$ (see Fig.~\ref{figure1}). 
In Fig.~\ref{figure5}(a) we show $R(\phi)$ results for the first quadrant of the electron pocket
around $[\pi,0]$ for 5 different values of doping $0.05 \leq \x \leq 0.4$.
In Fig.~\ref{figure5}(b) it is shown how the size of the $[\pi,0]$ electron pocket increases with electron doping
for the same $\x$ values as in panel (a).
Given the $C_4$ symmetry of ${\rm BiS_2}$, the pattern shown in Fig.~\ref{figure5}(a) repeats itself for all 
quadrants, with the appropriate $X \leftrightarrow Y$ swap in the definition of $R(\phi)$ \cite{note1}.
The results in Fig.~\ref{figure5}(a) show a decrease in mixing as the doping increases. 
Referring to Fig.~\ref{figure3}, it is easy to see that this is due to the increase 
in size of the electron pockets centered around the \emph{X} points in the BZ [see Fig.\ref{figure5}(b)]: 
as these pockets increase, larger parts of the Fermi surface will be in $R(\phi) \approx 0$ 
regions of the BZ. Keeping in mind, as discussed above, that ${\rm \Delta_{XY}}$ is associated  
to orbital-mixing ($R(\phi) \approx 1$) and ${\rm \Delta_{XX}}$ with the absence of it ($R(\phi) \approx 0$), 
one should then expect, as $\x$ varies, a maximum (at low doping) in ${\rm \Delta_{XY}}$ and a 
steady increase in ${\rm \Delta_{XX}}$. Indeed, if one takes a fixed $\eta = 0.4$, for example, 
in Fig.~\ref{figure2} (see dashed line in both panels), and contrasts how ${\rm \Delta_{XY}}$ and 
${\rm \Delta_{XX}}$ vary as $\x$ increases from zero, that is exactly what happens. 

\subsection{Orbital-mixing and multiband SC above the Lifshitz transition}\label{high}

Now, using the same ideas as in the previous section, 
we will explain the main structures of ${\rm \Delta_{XY}}$ and ${\rm \Delta_{XX}}$ at, and above, the Lifshitz 
transition. As mentioned above, as one approaches the Lifshitz transition (at $\x \approx 0.45$) 
from below (and the $[\pi,0]$ and $[0,\pi]$ 
electron pockets are about to touch and become hole pockets around $[0,0]$ and $[\pi,\pi]$) 
$R(\phi) \approx 0$ over the full extension of the Fermi surface, causing ${\rm \Delta_{XY}}$ to vanish. 
In reality, as the $R(\phi)$ results in Fig.~\ref{figure5} show, ${\rm \Delta_{XY}}$ should have 
essentially vanished for $\x \approx 0.2$, which agrees with the results in Fig.~\ref{figure2}. 
Figure \ref{figure6} explains the behavior of 
${\rm \Delta_{XY}}$ for $\x > 0.55$, where, as shown in Fig.~\ref{figure6}(b), an electron 
pocket around $[\pi,0]$ forms again and increases with $\x$. Fig.~\ref{figure6}(a) shows that 
this pocket initially presents strong orbital-mixing ($R(\phi) \approx 1.0$ for the whole pocket), 
which slowly decreases as the pocket increases, leading to the broad maximum 
in ${\rm \Delta_{XY}}$ around $\x \approx 0.6$, as seen in Fig.~\ref{figure2}(a). 

As to ${\rm \Delta_{XX}}$, at the Lifshitz transition it should reach a maximum, since, above it, 
the hole pockets $[0,0]$ and $[\pi,\pi]$ will start decreasing (as $\x$ keeps increasing).
This is shown in Fig.~\ref{figure7}(a) for $\x=0.46$ [(black) squares], $\x=0.6$ [(green) up triangle], 
and $\x=0.75$ [(cyan) diamonds]. The behavior of $R(\phi)$ for $\x=0.46$ and $\x=0.75$ is shown
in Fig.~\ref{figure7}(b) for the $[0,0]$ hole pocket and \ref{figure7}(c) for $[\pi,\pi]$. 
As expected, based on the results shown in Fig.~\ref{figure3}, the orbital-mixing ratio $R(\phi)$ is very small
for both pockets at $\x=0.46$ [(black] squares) and it barely changes between $\x=0.55$
(not shown) and $\x=0.75$ [(red) circles], indicating that these pockets only contribute to
${\rm \Delta_{XX}}$, as discussed above. Therefore, a somewhat broad maximum in
${\rm \Delta_{XX}}$, as shown in Fig.~\ref{figure2}(b), occurs around the Lifshitz transition at
$\x \approx 0.46$ and it is associated to the larger Fermi surface at this doping. 
Incidentally, the largest gap value in Fig.~\ref{figure2} is that for ${\rm \Delta_{XX}}$ right after the 
Lifshitz transition, when the $[0,0]$ and $[\pi,\pi]$ hole pockets have maximum size and basically 
no mixture. 

\subsection{Critical Temperature Results}\label{tc}

\begin{figure}
\centering
\includegraphics[width =3.8in]{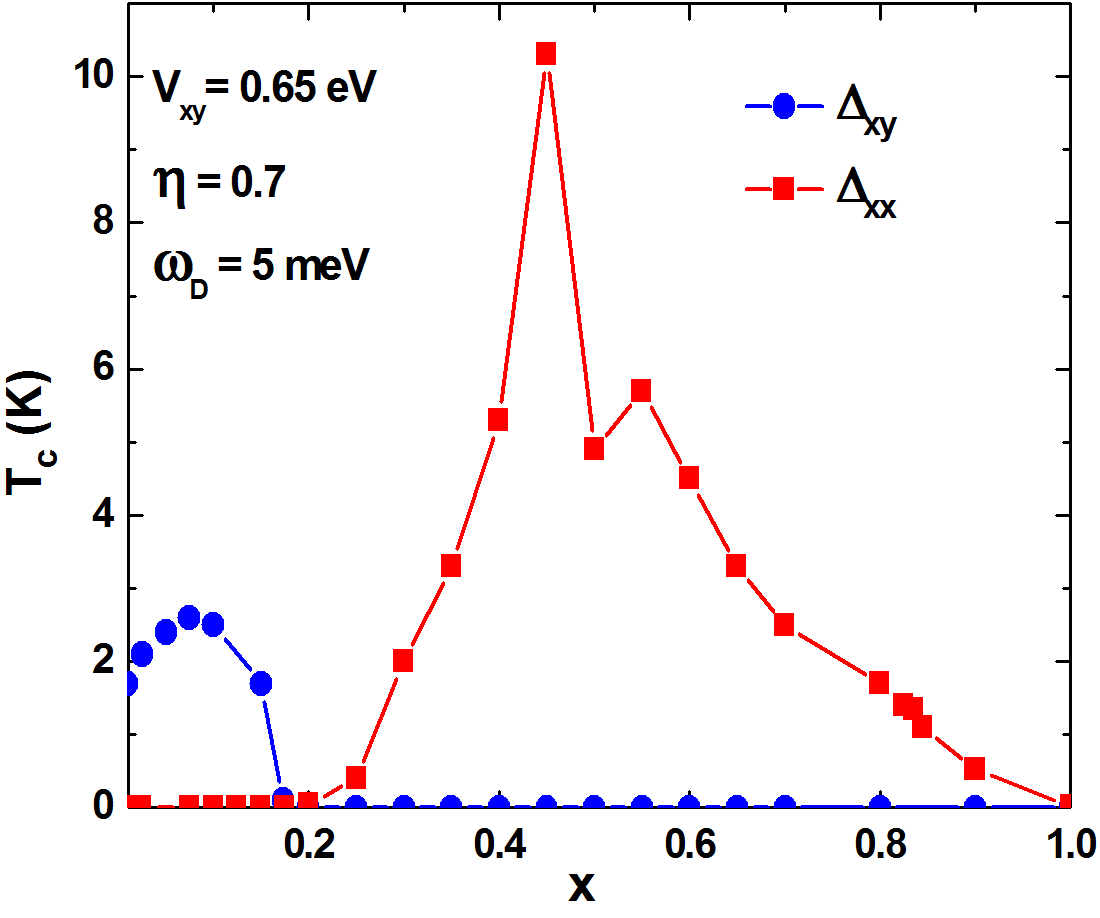}
	\caption{(Color online) Results for the superconducting critical temperature $T_c$ {\em vs} $\x$ obtained 
	from a plot very similar to that on Fig.~\ref{figure2}. The parameter values used here 
	were  $V_{XY}=0.16$~eV, $\omega_D=5$~meV, and ${\rm \eta=V_{XX}/V_{XY}}=0.7$. 
	A smaller value of $\omega_D$ than in Fig.~\ref{figure2} was used to insure 
	that the maximum $T_c$ value is close to that of $\rm BiS_2$ compounds \cite{Fang2015}.
}
\label{figure8}
\end{figure}

Figure \ref{figure8} shows results for the superconducting critical temperature $T_c$ {\em vs} $\x$ obtained from a 
plot (not shown) very similar to the one in Fig.~\ref{figure2}. The parameters used for calculating 
$T_c$ were $V_{XY}=0.16$~eV, $\omega_D=5$~meV, and ${\rm \eta=V_{XX}/V_{XY}}=0.7$. 
We used slightly smaller parameter values from the ones used in Fig.~\ref{figure2} 
so that the maximum $T_c \approx 10$~K is similar to that measured for ${\rm BiS_2}$ 
compounds \cite{Fang2015}. A comparison with a compilation of $T_c$ values for 
${\rm LnO_{1-x}F_xBiS_2}$ (where Ln = La, Ce, Pr, and Nd) [see Fig.~2 in Ref.~\cite{Fang2015}], 
shows an overall similarity with the results in Fig.~\ref{figure8}: $T_c \approx 2$~K for $\x =0.1$, 
$T_c \approx 6$~K for $\x \approx 0.6$, and (for Ln = Nd) a dip in $T_c$ for larger values of $\x$. 
Our results, therefore, are in good qualitative agreement with experimental results for ${\rm BiS_2}$ compounds. 
The $T_c$ values were obtained by solving, for a fixed value of $\x$, 
eqs.~\ref{D2new} and \ref{D3new} self-consistently for temperatures 
$T \geq 0$; then, a critical temperature associated to each diferent gap, ${\rm \Delta_{XY}}$ [(blue) circles] 
or ${\rm \Delta_{XX}}$ [(red) squares], is determined once the corresponding gap value falls below $10^{-6}$.

\section{Summary and Conclusions}\label{conc} 

In summary, we have presented results for a mean-field treatment of a model 
for multiband SC of ${\rm BiS_2}$-based layered compounds, using as starting point a minimal two-orbital tight-binding model, 
known to reproduce the main properties of the Fermi surface of ${\rm LaO_{(1-x)}F_xBiS_2}$, 
including its variation with electron doping. In this minimal model, the bands crossing the Fermi surface 
originate from the ${\rm 6p_X}$ and ${\rm 6p_Y}$ Bismuth orbitals (labeled ${\rm X}$- and ${\rm Y}$-orbital). 
The attractive pair-scattering part of the Hamiltonian allows for the formation of 
all three types of Cooper pairs (${\rm XX}$, ${\rm YY}$, and ${\rm XY}$), resulting in 
(after symmetry considerations) two gap equations involving superconducting order parameters ${\rm \Delta_{XX}}$ 
(associated to pair scatterings of the type ${\rm XX \leftrightarrow XX}$, ${\rm YY \leftrightarrow YY}$, 
and ${\rm XX \leftrightarrow YY}$) and ${\rm \Delta_{XY}}$ (${\rm XY \leftrightarrow XY}$). 
The self-consistent numerical solution of the gap equations was presented as a function 
of ${\rm \eta=V_{XX}/V_{XY}}$, the ratio between the pairing couplings, and the 
electron doping $\x$. We then defined the quantity $R(k_x,k_y)$, which measures the 
degree of ${\rm X}$- and ${\rm Y}$-orbital mixing of a band state, and used its value $0 \leq R(\phi) \leq 1$ 
to classify the Fermi surface pockets (parametrized through the polar angle $\phi$) as 
zero-mixing ($R(\phi) \approx 0$) or orbital-mixed ($R(\phi) \approx 1$). 
The definition of $R(\phi)$ allowed us to identify two 
distinct situations regarding SC: (i) zero-mixing pockets allow the formation of ${\rm XX}$ 
and ${\rm YY}$ pairs only, promoting ${\rm XX \leftrightarrow XX}$, ${\rm YY \leftrightarrow YY}$, 
and ${\rm XX \leftrightarrow YY}$ pair scattering, and therefore strengthens the ${\rm \Delta_{XX}}$ order parameter 
(or, as we call it, ${\rm XX}$-type SC), 
while (ii) orbital-mixed pockets result in ${\rm XY}$ pairs, promoting ${\rm XY \leftrightarrow XY}$ pair scattering 
and strengthening the ${\rm \Delta_{XY}}$ order parameter (${\rm XY}$-type SC). 
Calculating $R(k_x,k_y)$ in the first BZ we could 
assert that hole pockets around $[0,0]$ and $[\pi,\pi]$ are mostly zero-mixing and 
small electron pockets around  $[0,\pi]$ and $[\pi,0]$ are mostly orbital-mixed, becoming gradually zero-mixing as 
they increase in size. 
Based on that, and knowing how the Fermi surface pockets evolve with doping, we could semi-quantitatively predict 
the main structures observed in the ${\rm \Delta_{XY}}$ and ${\rm \Delta_{XX}}$ phase diagrams. 
In regions of the parameter space where both order parameters are present, we have, in general, that 
${\rm \Delta_{XX}} \gg {\rm \Delta_{XY}}$, unless $\eta \ll 1$.  This can be explained by the relative size of 
the areas in the first BZ where $R(k_x,k_y) \approx 0$ or $R(k_x,k_y) \approx 1$, with the former taking a 
much larger share of the first BZ. This implies that, unless the Fermi surface is restricted to small 
electron pockets around the \emph{X} points, which only occurs at very low doping, ${\rm XX}$-type SC will always 
dominate (unless $\eta \ll 1$). Finally, we also showed results for the superconducting critical temperature 
$T_c$, as a function of doping $\x$, which are in qualitative agreement with those measured for ${\rm BiS_2}$ compounds. 

In conclusion, in this work we present results for a particularly simple model of multiband SC, describing 
${\rm BiS_2}$ compounds, where the two bands crossing the Fermi surface originate from 
symmetry-related orbitals, and for which all types of Cooper pairs are allowed. The results 
for the two superconducting order parameters obtained can be semi-quantitatively linked 
to the way $R(\phi)$, the orbital-mixing ratio, changes as the Fermi surface evolves with 
electron doping. Given the current importance of multiband SC 
and the availability of computational techniques to produce 
effective models, at the tight-binding level, that describe the normal phase Fermi surface 
with relative accuracy, we envisage the use of the ideas here presented to 
spot favorable regions in the phase diagram (controlled mostly by carrier doping) to 
analyze specific aspects of multiband SC. For example, as clearly shown in this work, 
the use of the orbital-mixing ratio $R(\phi)$ allows us to correctly infer that ${\rm \Delta_{XY}}$ 
dominates at low electron doping, while ${\rm \Delta_{XX}}$ dominates close to the Lifshitz transition. 
This type of information could guide experimentalists into where to investigate for phenomena associated to each different 
order parameter. We hope that this approach would be appealing to experimental research groups 
interested in pinpointing favorable scenarios to observe such elusive phenomena as Leggett modes \cite{Lin2014}
or the Fulde-Ferrell-Larkin-Ovchinnikov state, \cite{Fulde1964,Larkin1965} which are associated to multiband SC. 
Finally, we also speculate that one could use 
these ideas to propose simple effective multiband models with the appropriate Fermi surfaces 
and orbital mixing, which will lead, at the appropriate band filling, to the dominance of one, or the other, 
of the superconducting order parameters. This could lead to proposals for real materials 
that could be described by these effective simple models, completing a reverse engineering strategy. 

\section*{Acknowledgements}
The authors thank MB Maple for comments about experimental results 
on ${\rm BiS_2}$ compounds. MAG and TOP acknowledge CNPq, MAC acknowledges CNPq and FAPERJ, 
and GBM acknowledges the Brazilian Government for financial support through a Pesquisador Visitante Especial 
grant from the Ci\^encias Sem Fronteiras Program, from the Minist\'erio da Ci\^encia, 
Tecnologia e Inova\c{c}\~ao. 


\bibliography{Bis2-experiment,Bis2-paper,Bis2-review,Bis2-theory,notes}

\begin{thebibliography}{10}
\bibitem{note2}
Note that, in solving eqs.~(13) and (14), the summation over $k$, following
  standard procedures, is performed around the Fermi surface using a cutoff
  energy $\omega=10$~meV.
\bibitem{note1}
Despite the fact that the pocket in Fig.~\ref{figure4}(b) is not centered
  around the origin of the Brillouin zone, it still holds true that points
  $\vec{k}$ and $-\vec{k}$ at the Fermi surface behave as shown in
  Fig.\ref{figure3}(a), i.e., the corresponding Bloch states have band-mixing
  ratio $R(\phi)$.
\end{thebibliography}

\bibliographystyle{iopart-num}

\end{document}